\documentclass[%
 reprint,
 amsmath,amssymb,
 aps,
showkeys
]{revtex4-1}
\usepackage{graphicx}% Include figure files
\usepackage{dcolumn}% Align table columns on decimal point
\usepackage{bm}% bold math
\usepackage{verbatim}
\usepackage{soul,xcolor}
\usepackage{amsmath,amstext}
\usepackage{amsmath,amsthm,amsfonts,amssymb}
\usepackage{mathtools,amssymb}
\usepackage[colorlinks=true]{hyperref}
\hypersetup{linkcolor = black, citecolor = black}

\let\Im\undefined

\DeclareMathOperator{\Im}{\mathrm{Im}}
\DeclareMathOperator\erf{erf}
\usepackage{hyperref}
\newcommand{\Kappa}{\mathcal{K}}

\newcommand{\calS}{\mathcal{S}}

\newcommand{\calT}{\mathcal{T}}
\newcommand{\calC}{\mathcal{C}}

\begin{document}

\preprint{APS/123-QED}

\title{Black String Thermodynamics in Noncommutative Spacetime: Anomaly and Phase Transition}

\author{Jo\~ao Chakrian}%

\author{Ant\^{o}nio de P\'{a}dua Santos}

\affiliation{%
 Departamento de F\'{i}sica, Universidade Federal Rural de Pernambuco, 52171-900 Recife, PE, Brazil
 }%


%\affiliation{
%\affiliation{
%\affiliation{%


\date{\today}

\begin{abstract}
Black holes have been a subject of investigation over years not only because they have interesting physical properties, but also because they seem to be the appropriate tool for studying gravity in quantum scale. Although a lot of effort has been made to understand the aspects of spacetime on the quantum scale, the approach which includes noncommutativity of spacetime proves to be promising among several possibilities. In this paper, we use the Hamilton-Jacobi method to study the thermodynamic properties of cylindrical black holes (black strings) in noncommutative spacetime. It is also investigated the behavior of the black string in the presence of back reaction as well as its influence on the system can be understood as an anomaly. This work aims to provide understanding about black strings thermodynamics in noncommutative spacetime since this scale of spacetime is important for the comprehension of the relation between quantum mechanics and general relativity.
\end{abstract}


\maketitle

%%%%%%%%% %%%%%%%%% %%%%%%%%% %%%%%%%%% %%%%%%%%% %%%%%%%%% %%%%%%%%% %%%%%%%%% %%%%%%%%% %%%%%%%%%
\section{\label{sec:intro} Introduction}
%%%%%%%%% %%%%%%%%% %%%%%%%%% %%%%%%%%% %%%%%%%%% %%%%%%%%% %%%%%%%%% %%%%%%%%% %%%%%%%%% %%%%%%%%%

The black hole solutions of the Einstein equations plays an important role in modern physics. Earlier, only the spherically symmetric black holes were studied because they are a great description of the final state of a gravitational collapse of stars. The interest of describing a gravitational collapse with different topology would not be continued so long due the formulation of hoop conjecture, established by Thorne \cite{Thorne}, which states that event horizons can be created only if a mass is compressed into a region with circumference less than $4\pi GM$ in every direction, which means that the cylindrical black holes should not exist. However, the hoop conjecture holds only when cosmological constant is assumed to be zero, then if one admits negative cosmological constant, the spacetime shall become asymptotically Anti-de Sitter \cite{Saifullah-thermo,Lemos-bs}. This fact allows to investigate cylindrically symmetric black hole solutions. 

The knowledge of spacetime in short scales has been a subject of great interest since it could be the key for a description of gravity in quantum world. Recently, Nicolini \cite{Nicolini1} presented a complete review about the relation between quantum black holes and the noncommutative geometry and one can observe the importance of this subject for the formulation of a quantum theory of gravity. In addition, the connection that string theory regards with noncommutative geometry has been a motivation for the attention given to this area, since the string theory and the quantum loop gravity are the main candidates for a theory of quantum gravity.

%%%%%%%%% %%%%%%%%% %%%%%%%%% %%%%%%%%% %%%%%%%%% %%%%%%%%% %%%%%%%%% %%%%%%%%% %%%%%%%%% %%%%%%%%%
\section{\label{sec:black_string} Black string}
%%%%%%%%% %%%%%%%%% %%%%%%%%% %%%%%%%%% %%%%%%%%% %%%%%%%%% %%%%%%%%% %%%%%%%%% %%%%%%%%% %%%%%%%%%

The papers \cite{Lemos-bs,Lemos-rotating-bs} performed by Lemos were a breakthrough in the subject of cylindrical black holes since he has shown that the cylindrical black holes, called black strings, are predicted when a negative cosmological constant is considered. Moreover, it is possible  to relate this (3+1)D solution to the BTZ (2+1)D solution \cite{Lemos-btz-bs}. Moreover, the BTZ solution has drawn attention to several geometric and topological properties such as those studied in \cite{Carvalho2003loop}. In such initial papers, the black string solution arises from solving the motion equations that result from the Einstein-Hilbert four dimensions action

\begin{equation} \label{eq:E-H_action}
S = \frac{1}{16\pi G}\int \sqrt{-\mathrm{g}}(R - 2\Lambda)d^4x,
\end{equation}
where $S$ is the Einstein-Hilbert action in four dimensions, $\textsl{g}$ is the determinant of the metric tensor, $\textsl{g}=\text{det}(g_{\mu\nu})$, $R$ is the Ricci scalar, $\Lambda$ is the cosmological constant and $G$ is the Newtonian gravitational constant. The black string solution is given by \cite{Lemos-rotating-bs}

\begin{eqnarray} \label{eq:BS_metric}
ds^2 &=& -\left(\alpha^2r^2 - \frac{4M}{\alpha r}\right)dt^2 + \left(\alpha^2r^2 - \frac{4M}{\alpha r}\right)^{-1}dr^2 \nonumber\\
&+& r^2d\phi^2 + \alpha^2r^2dz^2,
\end{eqnarray}
where $-\infty < t < \infty$, the radial coordinate $0\leq r < \infty$, the angular coordinate $0 \leq \phi < 2\pi$, the axial coordinate $-\infty < z < \infty$, $\alpha^2 = -\frac{\Lambda}{3} > 0$ and $M$ the mass per unit length in the $z$-direction. This solution represents a static straight black string with one event horizon at 

\begin{equation} \label{eq:horizonte_BS}
    r_+  = \frac{(4M_+)^{\frac{1}{3}}}{\alpha}
\end{equation}

%%%%%%%%% %%%%%%%%% %%%%%%%%% %%%%%%%%% %%%%%%%%% %%%%%%%%% %%%%%%%%% %%%%%%%%% %%%%%%%%% %%%%%%%%%
\section{\label{sec:NC_spacetime} Noncommutative spacetime}
%%%%%%%%% %%%%%%%%% %%%%%%%%% %%%%%%%%% %%%%%%%%% %%%%%%%%% %%%%%%%%% %%%%%%%%% %%%%%%%%% %%%%%%%%%

The study of noncommutative geometry, in context of physics, starts along with quantum mechanics and it was formal made by Snyder \cite{Snyder} in 1947. In noncommutative geometry, the coordinates of spacetime are replaced by operators. So we write \cite{Szabo}

\begin{equation}
x^\mu \longrightarrow \hat{x}^\mu
\end{equation}
that obeys

\begin{equation}
[\hat{x}^\mu,\hat{x}^\nu] = i\theta^{\mu\nu}
\end{equation}
where $\theta^{\mu\nu}$ is an anti symmetric matrix $D\times D$, $D$ is the dimension of spacetime and $\mu,\nu = 0,1,...,D-1$. Since the coordinates do not commute, one can write the generalized uncertain principle as

\begin{equation}
(\Delta x^\mu)^2(\Delta x^\nu)^2 \geq \left(\frac{1}{2i}[\hat{x}^\mu,\hat{x}^\nu]\right)^2,
\end{equation}
and after performing the calculations starting from the above equation, we find

\begin{equation}
\Delta x^\mu \Delta x^\nu \geq \frac{1}{2}\left|\theta^{\mu\nu}\right|.
\end{equation}
This relation shows that the point notion does not make sense anymore: if one knows precisely one coordinate, say $x^\mu$, then the other coordinate, say $x^\nu$, shall have a great uncertain. This fact implies that one cannot know both coordinates simultaneously. Therefore, it may be seen that a point cannot be defined since it is necessary to establish precisely a pair of coordinates to do so.

%%%%%%%%% %%%%%%%%% %%%%%%%%% %%%%%%%%%
\subsection{\label{subsec:NC_BS} Black string noncommutative spacetime} %\subsection{\label{subsec:matter_NC} Matter in noncommutative spacetime}
%%%%%%%%% %%%%%%%%% %%%%%%%%% %%%%%%%%%

Since the noncommutative spacetime does not admit the notion of point, we understand that the mass is no longer located, but it has a Gaussian distribution of length $\sqrt{\theta}$. Therefore, we can write the mass density $\rho_\theta$ as \cite{Nicolini1,Nicolini2,Singh}

\begin{equation} \label{eq:rho_theta}
\rho_\theta = \frac{M}{(4\pi\theta)^{\frac{3}{2}}} e^{-\frac{r^2}{4\theta}},
\end{equation}
that leads to

\begin{equation} \label{eq:m_theta}
m_\theta (r) = \frac{2M}{\sqrt{\pi}}\gamma\left(\frac{3}{2},\frac{r^2}{4\theta}\right)
\end{equation}
where $\gamma\left(\frac{3}{2},\frac{r^2}{4\theta}\right)$ is the incomplete lower gamma function. The reader may find more information about this function in the appendix.

%\subsection{\label{subsec:NC_BS} Black string noncommutative spacetime}

When one replaces the mass term in \eqref{eq:BS_metric} by the mass term \eqref{eq:m_theta}, one finds the metric that describes the noncommutative spacetime of a black string as follows

\begin{eqnarray} \label{eq:NC_BS_metric}
ds^2 &=& - \left[\alpha^2r^2 - \frac{8M}{\sqrt{\pi}\alpha r}\gamma\left(\frac{3}{2},\frac{r^2}{4\theta}\right)\right]dt^2 \nonumber\\
&+& \left[\alpha^2r^2 - \frac{8M}{\sqrt{\pi}\alpha r}\gamma\left(\frac{3}{2},\frac{r^2}{4\theta}\right)\right]^{-1}dr^2 \nonumber\\
&+& r^2 d\phi^2 + \alpha^2 r^2 dz^2.
\end{eqnarray}
and we can find the event horizon, using the vanishing of the component of the metric tensor $\textsl{g}_{tt} = 0$, as follows

\begin{equation} \label{eq:NC_BS_event horizon}
r_+^3 = \frac{8M_+}{\sqrt{\pi}\alpha^3}\gamma\left(\frac{3}{2},\frac{r_+^2}{4\theta}\right).
\end{equation}

%%%%%%%%% %%%%%%%%% %%%%%%%%% %%%%%%%%% %%%%%%%%% %%%%%%%%% %%%%%%%%% %%%%%%%%% %%%%%%%%% %%%%%%%%%
\section{\label{sec:Tunneling_NC_BS} Quantum tunneling from scalar field in black string noncommutative spacetime}
%%%%%%%%% %%%%%%%%% %%%%%%%%% %%%%%%%%% %%%%%%%%% %%%%%%%%% %%%%%%%%% %%%%%%%%% %%%%%%%%% %%%%%%%%%

The approach that we use is to perform the Hamilton-Jacobi method to describe the tunneling on the event horizon of the black string. This method is based on the tunneling approach by Parikh and Wilczeck \cite{Parikh}, the path integrals method by Srinivasan \cite{Srinivasan-complex} and can be found in references \cite{Saifullah-emission} and \cite{Saifullah-quantum}. We start from the Klein-Gordon equation on curved spaces for a scalar field $\Phi$

\begin{equation} \label{eq:K-G}
\mathrm{g}^{\mu\nu}\partial_\mu\partial_\nu\Phi - \frac{m^2}{\hbar^2}\Phi = 0,
\end{equation}
where $\mathrm{g}^{\mu\nu}$ is the metric tensor related to the line element \eqref{eq:NC_BS_metric} and $m$ is the mass of the field. In order to solve this equation, we use the ansatz

\begin{equation} \label{eq:WKB}
\Phi(t,r,\phi,z) = \exp\left\{\frac{i}{\hbar}S(t,r,\phi,z)\right\},
\end{equation}
where $S$ represents the trajectory action associated to the particle tunneling. We can recognize this solution as the WKB approximation \cite{Birrell}. This low wavelength solution is justified since the amount of null geodesics tends to infinity at the event horizon which means that there is a blueshift \cite{Parikh}. Substituting \eqref{eq:WKB} into \eqref{eq:K-G} and using the first term of the expansion of the action in terms of $\hbar$, we find

\begin{equation}
\mathrm{g}^{tt}(\partial_t S)^2 + \mathrm{g}^{rr}(\partial_r S)^2 + \mathrm{g}^{\phi\phi}(\partial_\phi S)^2 + \mathrm{g}^{zz}(\partial_z S)^2 + m^2 = 0.
\label{eq:action_part}
\end{equation}
The solution to equation \eqref{eq:action_part} can be considered by separation of variables as follows
\begin{equation}
S(t,r,\phi,z) = - Et + W(r) + J_\phi \phi + J_z z + C,
\end{equation}
where $E$ is a constant associated to the energy, $W(r)$ is a function to be determined, $J_\phi,J_z$ are constants associated to angular momentum in $\phi$ and $z$ direction, respectively, and $C$ is a constant. After performing an integration on complex plane, the function $W(r)$ is given by

\begin{eqnarray}\label{eq:W_+/-}
W_{\pm}(r) &=& \pm \pi iE\left\{2\alpha^2r_+ + \frac{8M_+}{\sqrt{\pi}\alpha}\left[\frac{1}{r^2_+}\gamma\left(\frac{3}{2},\frac{r^2_+}{4\theta}\right)\right.\right.\nonumber\\
&-& \left.\left.  - \frac{1}{r_+}\gamma'\left(\frac{3}{2},\frac{r^2_+}{4\theta}\right)\right]\right\}^{-1},
\end{eqnarray}
where the prime indicates derivative with respect to $r_+$. The probabilities of crossing the event horizon are given by

\begin{eqnarray}
\Gamma_{\mbox{\footnotesize emission}} &=&  \exp\left\{-\frac{2}{\hbar}[\Im(W_+) + \Im(C)]\right\} \label{eq:Gamma_emission1}\\
\Gamma_{\mbox{\footnotesize absorption}} &=& \exp\left\{-\frac{2}{\hbar}[\Im(W_-) + \Im(C)]\right\} \label{eq:Gamma_absorption},
\end{eqnarray}
and by the condition that the probability of entering in the event horizon of black string is $100\%$ \cite{Saifullah-emission}, we determine that $\Im(W_-) = \Im(C)$ and the replacement of this result into \eqref{eq:Gamma_emission1} leads to

\begin{widetext}
\begin{eqnarray} \label{eq:Gamma_emission}
\Gamma_{\mbox{\footnotesize emission}} &=& \exp\left\{- \frac{4}{\hbar}\Im(W_+)\right\} \nonumber\\
&=& \exp\left\{-\frac{\pi E}{\hbar}\left[\frac{\alpha^2r_+}{2} + \frac{2M_+}{\sqrt{\pi}\alpha}\left[\frac{1}{r^2_+}\gamma\left(\frac{3}{2},\frac{r^2_+}{4\theta}\right) - \frac{1}{r_+}\gamma'\left(\frac{3}{2},\frac{r^2_+}{4\theta}\right)\right]\right]^{-1}\right\}.
\end{eqnarray}
\end{widetext}
Then, if one compares the emission factor \eqref{eq:Gamma_emission} above to the canonical ensemble Boltzmann factor $e^{-\frac{E}{T_H}}$, the Hawking temperature is

\begin{equation} \label{eq:T_H_NC}
T_H = \frac{\alpha^2 r_+}{4\pi}\left[3 - \frac{r_+^3}{4\theta^{\frac{3}{2}}} \frac{e^{-\frac{r_+^2}{4\theta}}}{\gamma\left(\frac{3}{2},\frac{r_+^2}{4\theta}\right)}\right].
\end{equation}
where we consider $E > 0$, $\hbar = 1$, and the equation \eqref{eq:NC_BS_event horizon} for writing $M_+$ in terms of $r_+$.
An equivalent expression was obtained by \cite{Singh} performing a different method. If we take the commutative spacetime limit $\frac{r_+}{\sqrt{\theta}} \rightarrow \infty$ we obtain 

\begin{equation} \label{eq:T_H_C}
T_H^{(c)} = \frac{3\alpha^2}{4\pi}r_+.
\end{equation}
This result was obtained by the references \cite{Saifullah-thermo} and \cite{Cai}. We notice that the equation \eqref{eq:T_H_NC} recovers the known behaviour in commutative spacetime when we do not take account the effects of the noncommutativity. We can compare the result \eqref{eq:T_H_NC} with the temperature in the spherical case given by \cite{Nicolini1}. We note from equation (\ref{fig:tempNC_tempC}) that in cylindrically symmetric case, the temperature does not diverge when near to the origin and the curve does not cross the horizontal axis as it happens in spherically symmetric case. 

%We understand that it is due to the fact that the black string in noncommutative spacetime has only one event horizon whereas the noncommutative Schwarzchild black hole has more than one [Ref-Banerjee].

%%%%%%%%% %%%%%%%%% %%%%%%%%% %%%%%%%%% %%%%%%%%% %%%%%%%%% %%%%%%%%% %%%%%%%%% %%%%%%%%% %%%%%%%%%
\section{\label{sec:termo} Thermodynamics}
%%%%%%%%% %%%%%%%%% %%%%%%%%% %%%%%%%%% %%%%%%%%% %%%%%%%%% %%%%%%%%% %%%%%%%%% %%%%%%%%% %%%%%%%%%

Since the cylindrical black hole can be understand as a thermodynamic system, we interpret the results in a thermodynamic point of view. The entropy can be determined by starting from the first law for black holes

\begin{equation*}
dM_+ = T_H dS_+
\end{equation*}
which means that the entropy can be obtained after performing the integration

\begin{equation} \label{eq:entropy_variation}
\Delta S_+ = \int \frac{dM_+}{T_H} = \int_{r_0}^{r_+} \frac{1}{T_H}\left(\frac{\partial M_+}{\partial r'_+}\right)dr'_+
\end{equation}
and if one sets $r_0 = 0$, the entropy results

\begin{equation} \label{eq:entropy}
S_+ = \frac{\pi^{\frac{3}{2}}\alpha}{2}\int_0^{r_+}\frac{r'_+}{\gamma\left(\frac{3}{2},\frac{{r'_+}^2}{4\theta}\right)}dr'_+,
\end{equation}
this expression was obtained before by \cite{Singh} and represents the entropy of the cylindrical black hole in noncommutative spacetime. It can be evaluated by performing a numerical integration or an asymptotic expansion of the function $\gamma$  in the commutative spacetime limit $r_+/\sqrt{\theta}$. Taking the commutative spacetime limit in the equation \eqref{eq:entropy}, we obtain

\begin{equation} \label{eq:entropy_C}
S_+^{(c)} = \frac{\pi^{\frac{3}{2}}\alpha}{2}\int_0^{r_+} \frac{r'_+}{\sqrt{\pi}/2}dr'_+ = \frac{\pi\alpha r_+^2}{2},
\end{equation}
where we have used the fact that $\gamma(a,x\rightarrow\infty) = \Gamma(a)$ (as equation \ref{eq:expan_gama_1}). Note that the above expression obeys the Bekenstein area law $S = A/4$ since the area of the black string is given by $A = 2\pi\alpha r_+^2$ \cite{Cai}.

\begin{figure}
	%\vspace{0.5cm}
	\includegraphics[scale=.48]{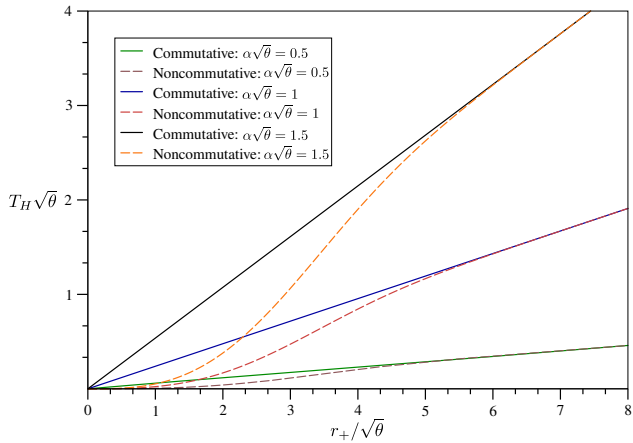}
	\caption{\label{fig:tempNC_tempC}\footnotesize Temperature. The horizontal axis is normalized by $\sqrt{\theta}$ and the vertical axis is multiplied by this factor. We used numerical values for the parameter $\alpha\sqrt{\theta}$ in each plot.}
	%\vspace{-0.5cm}	
	\end{figure}

Observing the figure \ref{fig:tempNC_tempC} we see that noncommutativity affects the temperature and, in the scale $r_+ \approx 5\sqrt{\theta}$, turns the cylindrical black hole colder than the expected in only commutative prediction. By investigating figure \ref{fig:tempNC_tempC} we notice that to reach the $T_H = 0$, it would be necessary to get $r_+ = 0$, such fact is not permitted by hypothesis since the spacetime has a minimal length $\sqrt{\theta}$. 
	
In order to investigate the stability of the system, we calculate the heat capacity of the cylindrical black hole. Considering the first law for black holes $dM_+ = T_HdS_+$ we can write a relation for the heat capacity as

\begin{equation*}
C_+ = \frac{\partial Q}{\partial T_H} = \frac{T_H\partial S_+}{\partial T_H} = \frac{\partial M_+}{\partial T_H} = \frac{\left(\frac{\partial M_+}{\partial r_+}\right)}{\left(\frac{\partial T_H}{\partial r_+}\right)},
\end{equation*}
which can be calculated using the relation \eqref{eq:T_H_NC} for the temperature in terms of $r_+$ and by the expression of mass that can be obtained by simple manipulating the equation \eqref{eq:NC_BS_event horizon}:

\begin{equation} \label{eq:M_+}
M_+ = \frac{\sqrt{\pi}\alpha^3}{8}\frac{r_+^3}{\gamma\left(\frac{3}{2},\frac{r_+^2}{4\theta}\right)}.
\end{equation}
The heat capacity is given by 

\begin{equation} \label{eq:C_NC}
C_+ = \frac{\pi^{\frac{3}{2}}}{2}\alpha r_+^2 \frac{\left(3 - \frac{r_+^3}{4\theta^{\frac{3}{2}}}e^{-\frac{r_+^2}{4\theta}}\right)}{\left[3\gamma - r_+\left(\gamma'' r_+ - \frac{{\gamma'}^2}{\gamma}r_+ + 2\gamma'\right)\right]},
\end{equation}
where $\gamma \equiv \gamma\left(\frac{3}{2},\frac{r_+^2}{4\theta}\right)$, and [$'$] indicates derivatives with respect to $r_+$. This result is equivalent to the result obtained in reference \cite{Singh}. In the commutative limit $r_+/\sqrt{\theta} \rightarrow \infty$ one recovers the result

\begin{equation} \label{eq:C_+_comut}
    C_+^{(c)} = \pi\alpha r_+^2
\end{equation}

\begin{figure}
	\includegraphics[scale=.48]{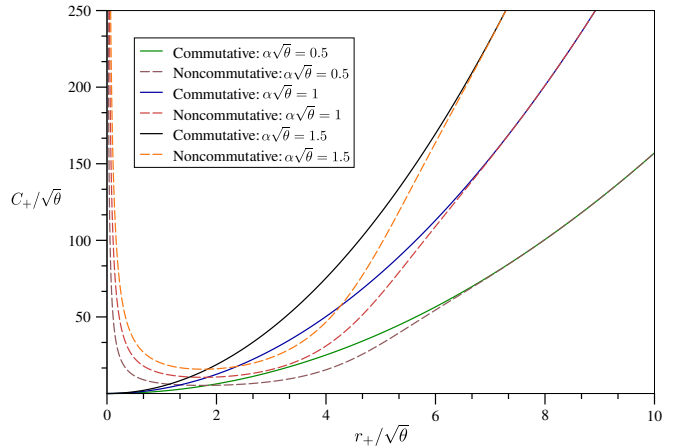}
	\caption{\label{fig:capaNC_capaC} Heat capacity. The horizontal axis is normalized by $\sqrt{\theta}$ as well as the vertical axis. We used numerical values for the parameter $\alpha\sqrt{\theta}$ in each plot.}
	\end{figure}

We plot these results in figure \ref{fig:capaNC_capaC}. One can note that the heat capacity is always positive $C_+ > 0$ in both commutative and noncommutative spacetimes, which means that the system is stable and phase transitions are not foreseen. It can be observed that the divergent behavior of the dashed curves nearby the origin means that it is necessary to transfer an infinity amount of energy to the system in order to reach $r_+ = 0$. For what we know, the heat capacity is related to the energy that should be transferred to the system to change its temperature.  We can also see that the temperature is related to the event horizon through the equation \eqref{eq:T_H_NC}. The figure \ref{fig:capaNC_capaC} shows that it would be necessary to transfer an infinity amount of energy to reach $T_H = 0$. In other words, this result is a verification of the third law of thermodynamics since it is not possible to reach the absolute zero temperature. On another hand, $r_+ \rightarrow 0$ when $T_H \rightarrow 0$ which means that to reach the zero temperature demands a complete evaporation. That is, the cylindrical black hole does not evaporate completely. This fact can be verified by checking the figure \ref{fig:massNC_massC} that shows the mass in terms of $r_+$. For each dashed curve it is possible to obtain a minimal value of mass when $r_+/\sqrt{\theta} \rightarrow 0$. Which one of dashed curves in figure \ref{fig:massNC_massC} is constructed by setting some value to the parameter $\alpha\sqrt{\theta}$ (i.e. $0.75, 1.25, 1.5$).

    \begin{figure}
    \centering
    \includegraphics[scale=0.5]{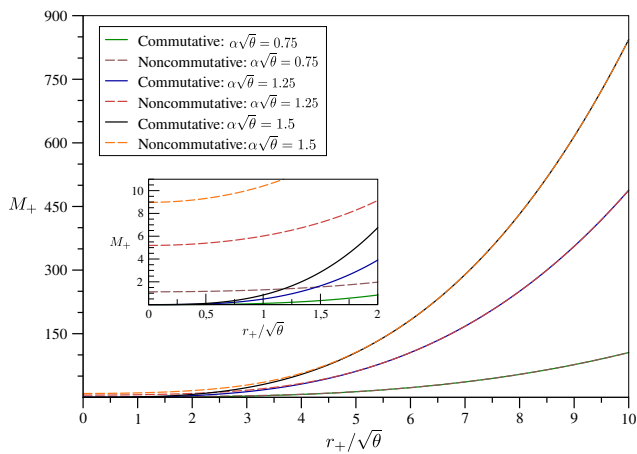}
    \caption{\label{fig:massNC_massC}\footnotesize Mass per unit of length. We can write the minimal values of mass to each noncommutative case: $M_{\mbox{\scriptsize min}} = 1.12$ for $\alpha\sqrt{\theta} = 0.75$; $M_{\mbox{\scriptsize min}} = 5.19$ for $\alpha\sqrt{\theta} = 1.25$; $M_{\mbox{\scriptsize min}} = 8.97$ for $\alpha\sqrt{\theta} = 1.5$.}
    \label{fig:massNC_massC}
    \end{figure}

It can be noticed that the thermodynamic quantities are greater as long as the parameter $\alpha\sqrt{\theta}$ is greater. Investigating figure \ref{fig:tempNC_tempC} it can be seen that the greater values of temperature are reached by the curves constructed by setting the parameter greater. This fact can be noticed by checking the figure \ref{fig:capaNC_capaC} for the heat capacity, and the figure  \ref{fig:massNC_massC} for the mass per unit of length. Such parameter defines the scale of the cosmological constant $\Lambda$ in terms of the minimal length $\sqrt{\theta}$ since $\alpha^2 = -\Lambda/3$. Considering the cosmological constant to be related to the acceleration of the universe expansion, as greater value is chosen to this acceleration as greater are the response by the thermodynamic quantities. %These observations can be extended to next quantities discussed such as the internal energy presented in figure \ref{fig:EintNC_EintC}. 

In order to calculate the free energy of the system, we first calculate the internal energy. This quantity shall be calculated following the method presented in \cite{Kim}: we define a cylindrical cavity of finite radius $R$ where $R > r_+$. The temperature outside the event horizon is written as

\begin{equation} \label{eq:T_cavity}
T = \frac{T_H}{\sqrt{\mathrm{g}_{tt}(R)}},
\end{equation}
where $\mathrm{g}_{tt}$ is the first component ($\mathrm{g}_{00}$) of the metric tensor given in \eqref{eq:BS_metric}. We see that the temperature term in equation \eqref{eq:T_cavity} is that one at event horizon $T_H$ weighted by the Tolman factor \cite{Kim}, which considers the redshift once we get further from the event horizon. From the first law for black holes $dE = TdS$, we write

\begin{equation} \label{eq:internal_energy_int}
E = M_0 + \int_{S_0}^{S_+} TdS'_+ = M_0 + \int_{r_0}^{r_+} T\left(\frac{\partial S'_+}{\partial r'_+}\right)dr'_+,
\end{equation}
where $r_0$ and $M_0$ are the initial values of $r_+$ and the mass, respectively. Here we assume that the energy of the black string is associated to its mass, and the radiation emitted implies a variation of the mass and the internal energy. If one writes the relations \eqref{eq:T_cavity} modified by \eqref{eq:BS_metric} and uses the equation \eqref{eq:entropy} to obtain  $\partial S'_+/\partial r'_+$, the internal energy in equation \eqref{eq:internal_energy_int} can be expressed as

\begin{widetext}
\begin{equation} \label{eq:internal_energy}
E = M_0 + \frac{\sqrt{\pi}}{\gamma\left(\frac{3}{2},\frac{R^2}{4\theta}\right)}\frac{\alpha^2R^2}{4}\left[\left(1 - \frac{\gamma\left(\frac{3}{2},\frac{R^2}{4\theta}\right)}{\gamma\left(\frac{3}{2},\frac{r_0^2}{4\theta}\right)}\frac{r_0^3}{R^3}\right)^{\frac{1}{2}} - \left(1 - \frac{\gamma\left(\frac{3}{2},\frac{R^2}{4\theta}\right)}{\gamma\left(\frac{3}{2},\frac{r_+^2}{4\theta}\right)}\frac{r_+^3}{R^3}\right)^{\frac{1}{2}}\right].
\end{equation}
\end{widetext}
Considering the commutative limit $r_+/\sqrt{\theta} \rightarrow \infty$ we obtain 

\begin{equation} \label{eq:internal_energy_C}
E^{(c)} =  M_0 + \frac{\alpha^2R^2}{2}\left[\left(1 - \frac{r_0^3}{R^3}\right)^{\frac{1}{2}} - \left(1 - \frac{r_+^3}{R^3}\right)^{\frac{1}{2}}\right].
\end{equation}

    \begin{figure}
	\includegraphics[scale=.56]{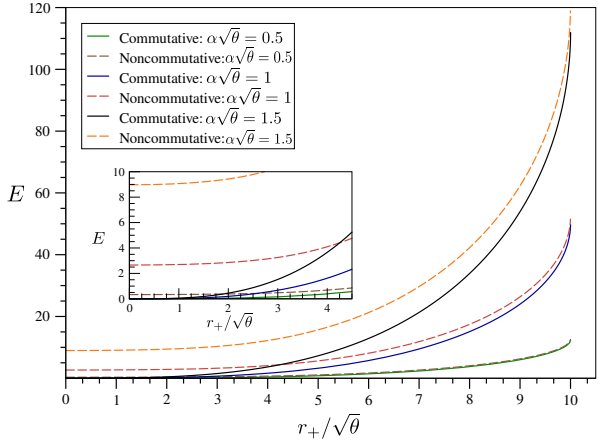}
	\caption{\label{fig:EintNC_EintC}\footnotesize Internal energy. Minimal values in noncommutative case: $E_{\mbox{\scriptsize min}} = 0.33$ for $\alpha\sqrt{\theta} = 0.5$; $E_{\mbox{\scriptsize min}} = 2.66$ for $\alpha\sqrt{\theta} = 1$; $E_{\mbox{\scriptsize min}} = 8.97$ for $\alpha\sqrt{\theta} = 1.5$. It was made $E_0 = 0$ and $r_0 = 0$.}
	\end{figure}

A plot with different parameters is presented in figure \ref{fig:EintNC_EintC}. It is possible to see that the dashed curves, which represent the internal energy in noncommutative spacetime, do not go to zero and there is a minimal value of energy in each case. One notices that as greater the parameter $\alpha\sqrt{\theta}$, greater are the values of internal energy reached, such fact corroborates with was earlier discussed about the response by the thermodynamic quantities. 

After the presentation of the equations \eqref{eq:internal_energy} and \eqref{eq:internal_energy_C}, we are able to discuss about the free energy associated to the system. We start from the thermodynamic relation $F = E - TS$ which in our work is better expressed as

\begin{equation} \label{eq:F_def}
F = E - TS_+.
\end{equation}
Realize that this equation cannot be easily solved since $S_+$ is defined by the integral in \eqref{eq:entropy}. We can avoid these difficulties by performing the asymptotic expansion of the incomplete lower gamma function, for large $x$, given as (as equation \ref{eq:expan_assin_gama}) %\cite{Banerjee}

\begin{eqnarray} \label{eq:gamma_expansion_l=0}
\gamma\left(\frac{3}{2},x\right) &\approx& \frac{\sqrt{\pi}}{2}\left[1 - e^{-x}\sum_{l=0}^{\infty} \frac{x^{\frac{(1-2l)}{2}}}{\Gamma\left(\frac{3}{2} - l\right)}\right] \nonumber\\
&=& \frac{\sqrt{\pi}}{2}\left[1 - \frac{x^{\frac{1}{2}}}{\sqrt{\pi}/2}e^{-x}\right],
\end{eqnarray}
where we have used the expansion in order $l = 0$, the first term. It is important to realize that the expression describing the free energy that shall be obtained is an approximated result: to take into account only the first term of the expansion \eqref{eq:gamma_expansion_l=0} implies that the other terms cannot contribute to the final result. Although this is the only approximation made in the entire calculation and the approximated result could be considered satisfactory. 

\begin{figure}[h]
	\includegraphics[scale=.49]{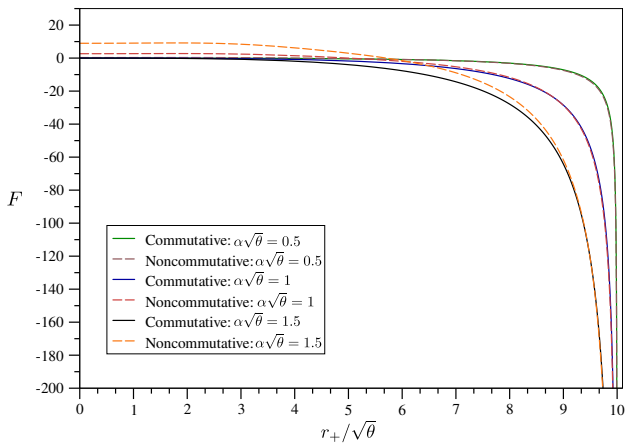}
	\caption{\label{fig:FNC_FC}\footnotesize  Free energy. Note that the sign of $F$ changes and the corresponding values in horizontal axis, in terms of $r_+/\sqrt{\theta}$, for commutative cases (straight curves) are: $0.05$ for $\alpha\sqrt{\theta} = 0.5$, $0.04$ for $\alpha\sqrt{\theta} = 1$, $0.03$ for $\alpha\sqrt{\theta} = 1.5$; for noncommutative cases (dashed curves): $4.07$ for $\alpha\sqrt{\theta} = 0.5$,  $4.99$ for $\alpha\sqrt{\theta} = 1$, and $5.7$ for $\alpha\sqrt{\theta} = 1.5$.}
	\end{figure}

Replacing both equations \eqref{eq:entropy} modified by the expansion \eqref{eq:gamma_expansion_l=0}, the equation \eqref{eq:internal_energy} into the relation \eqref{eq:F_def}, and substituting the expression of the temperature \eqref{eq:T_cavity}, after performing some calculations, the free energy is is expressed in figure \ref{fig:FNC_FC}.

\begin{comment}
given by 

\begin{widetext}
\begin{eqnarray} \label{eq:Free_energy}
F &=& M_0 + \frac{\sqrt{\pi}}{\gamma\left(\frac{3}{2},\frac{R^2}{4\theta}\right)}\frac{\alpha^2R^2}{4}\left[\left(1 - \frac{\gamma\left(\frac{3}{2},\frac{R^2}{4\theta}\right)}{\gamma\left(\frac{3}{2},\frac{r_0^2}{4\theta}\right)}\frac{r_0^3}{R^3}\right)^{\frac{1}{2}} - \left(1 - \frac{\gamma\left(\frac{3}{2},\frac{R^2}{4\theta}\right)}{\gamma\left(\frac{3}{2},\frac{r_+^2}{4\theta}\right)}\frac{r_+^3}{R^3}\right)^{\frac{1}{2}}\right] \nonumber\\
&-& \frac{\sqrt{\pi}\alpha^2}{\left[1 - \frac{\gamma\left(\frac{3}{2},\frac{R^2}{4\theta}\right)}{\gamma\left(\frac{3}{2},\frac{r_+^2}{4\theta}\right)}\frac{r_+^3}{R^3}\right]^{\frac{1}{2}}}\left[3 - \frac{r_+^3}{4\theta^{\frac{3}{2}}}\frac{e^{-\frac{r_+^2}{4\theta}}}{\gamma\left(\frac{3}{2},\frac{r_+^2}{4\theta}\right)}\right]\left[\frac{r_+^3}{8\sqrt{\pi}R} + \frac{\theta}{2\sqrt{\pi}}\frac{r_+}{R}\erf\left(\frac{r_+}{2\sqrt{\theta}}\right) + \left(\frac{6\theta + r_+^2}{12\pi\sqrt{\theta}}\right)\frac{r_+^2}{R}e^{-\frac{r_+^2}{4\theta}}\right]. 
\end{eqnarray}
\end{widetext}
Here, the functions $\gamma\left(\frac{3}{2},x\right)$ represents, as always, the gamma function of the argument $x$, and $\erf{x}$ represents the error function as presented in appendix.

\begin{equation} \label{eq:error_function}
\erf{x} = \int \frac{2e^{-x^2}}{\sqrt{\pi}} dx.
\end{equation}

The result \eqref{eq:Free_energy}  
\end{comment}

The figure \ref{fig:FNC_FC} shows that the free energy assumes both negative and positive values, this can be understood from investigating figure \ref{fig:tempNC_tempC} and equation \eqref{eq:F_def} because as greater $r_+$, greater is $T_H$ and the term $T_HS_+$ will be greater too. This implies that for some value of $r_+$, the term $T_HS_+$ will be greater than $E$ and the free energy $F = E - T_HS_+$ shall adopt negative values. Figure \ref{fig:FNC_FC} also shows that the curves concerning to the commutative case cross the curves that represent the noncommutative case before the overlap in the region $r_+/\sqrt{\theta} = 10$, where the noncommutativity does not exercise significant influence over the system. That is, for a determined radius, the noncommutative geometry reproduces the commutative case.  We see that the free energy, checking the commutative curves, has a similar behavior to each other even if the parameter $\alpha\sqrt{\theta}$ increases, and the main changes can be said to be the increase in the value of the sign change of $F$. Checking now the dashed curves, concerning to the noncommutative spacetime, we see that the values in the vertical axis increases whereas the numerical parameter mentioned increases. The increase in the parameter associated to the acceleration of the universe expansion is manifested by the increase in the values of $F$. The plot we are analyzing can be related to the figure \ref{fig:EintNC_EintC} and the figure \ref{fig:tempNC_tempC} if one investigates in which scale the free energy assumes negative values and consider the equation \eqref{eq:F_def}. Firstly, checking the figure \ref{fig:FNC_FC} we can realize that dashed curves assume values greater than the values assumed by the straight curves. Since the free energy depends on the positive value of internal energy and also it depends on the negative value of temperature multiplied by entropy. The difference between these quantities is positive and greater in noncommutative cases represented by the dashed curves for small values of $r_+$. This is a consequence of the internal energy behavior illustrated in the figure \ref{fig:EintNC_EintC} in which the dashed curves are above the straight curves. For small $r_+$, the figure \ref{fig:tempNC_tempC} presents that the temperature values performed by dashed curves are small which implies that the product $TS_+$ shall be small too. Therefore, $F$ should be greater. The large $r_+$ scale causes the difference in equation \eqref{eq:F_def} to decrease until the free energy to assume negative values.

%This fact corroborates to the interpretation, concerning to the plot \ref{fig:EintNC_EintC}, that the internal energy is greater when we take the noncommutativity into account.  

From the point of view of thermodynamics, it is known that the free energy can be related to the work done on or by the system in a reversible isothermal process. In our case, we can study the free energy by analyzing two different regions: the region which $F > 0$ and the other one which $F < 0$ and the points which determine the regions to each case can be seen in figure \ref{fig:FNC_FC}. We can write 

\begin{equation}
    \Delta F = \Delta E - T_H\Delta S_+.
\end{equation}
Considering the variation of internal energy

\begin{equation}
    \Delta E = T_H\Delta S_+ - W.
\end{equation}
Where $W$ represents work. It results that

\begin{equation} \label{eq:Delta_F}
    \Delta F = - W,
\end{equation}
that is, the variation of the free energy $F$ represents the work made by the system during the isothermal process and the sign of $\Delta F$ depends on the sign of $W$. In the following analysis, we use $\Delta F = F_f - F_i$ where the $F_f$ means final free energy of the process and $F_i$ means initial free energy of the process. In the region which $F > 0$, it always shall be $F_f > 0$ and $F_i > 0$ it means that if $F_f > F_i$ then $\Delta F > 0$ and the relation \eqref{eq:Delta_F} shows us that the work was made on the system and if $F_f < F_i$, then $\Delta F < 0$, the situation is inverted and work was made by the system. In the region which $F < 0$, if we have $|F_f| > |F_i|$ then $\Delta F < 0$ [because $- |F_f| - (-|F_i|) < 0$ since $|F_f| > |F_i|$] which means work done by the system and if $|F_f| < |F_i|$ then $\Delta F > 0$ and work was done on the system. Variations of $F$ between both regions can be analyzed checking in which region are located the initial and the final values of the free energy.

%%%%%%%%% %%%%%%%%% %%%%%%%%% %%%%%%%%% %%%%%%%%% %%%%%%%%% %%%%%%%%% %%%%%%%%% %%%%%%%%% %%%%%%%%%
\section{\label{sec:back} Back Reaction}
%%%%%%%%% %%%%%%%%% %%%%%%%%% %%%%%%%%% %%%%%%%%% %%%%%%%%% %%%%%%%%% %%%%%%%%% %%%%%%%%% %%%%%%%%%

Intuitively, the back reaction effect can be understood as the fact that gravity can act on itself, or more specifically, how the gravitational field influences itself, which makes Einstein's equations nonlinear \cite{Wald, Carroll}. This characteristic of non-linearity of the equations can be assimilated from an analysis of the Feynmann diagrams for gravitational interactions. From the Feynmann approach we can understand that the gravitational interaction is performed by the exchange of a virtual graviton (quantized metric field) and since gravitons themselves can exchange virtual gravitons, they can consequently exercise gravitational interaction on themselves \cite{Carroll,Schweber}. The consideration of the back reaction effect aims to provide a better approximation for the geometry of space-time on a quantum scale. The interest in such a study dates back to the fact that there is no consistent theory of quantum gravity, that is, the theory of gravitation in the scale which quantum effects influence spacetime. The first question concerning to the back reaction is how it is considered. The answer is given by semiclassical approach which couples a quantized matter field to the gravitational field using the semiclassical Einstein equations \cite{Ford,Wald}

\begin{equation} \label{eq:Einstein_semiclassica}
G_{\mu\nu} + \Lambda g_{\mu\nu} = 8\pi\langle T_{\mu\nu}\rangle^{(\mbox{\scriptsize ren})}.
\end{equation}
The equation \eqref{eq:Einstein_semiclassica} means that the spacetime geometry generates a non-zero vacuum expectation value of the energy-momentum tensor $\langle T_{\mu\nu}\rangle^{(\mbox{\scriptsize ren})}$ which in turn acts as a source of curvature. According to reference \cite{Lousto}, this is the back reaction problem. The upper index $\mbox{ren}$ means renormalized. The renormalization is necessary due the divergences in the calculations since the energy-momentum tensor is nonlinear. To avoid such problems, we try to answer the second question of our discussion concerning to the back reaction approach in curved spacetimes. We set $G = c = k_B = 1$, back reaction can be interpreted as a perturbation in the surface gravity and can be written as \cite{Banerjee}

    \begin{equation} \label{eq:kappa_perturb_def}
    \Kappa = \Kappa_0 (r_+) + \xi\Kappa_0(r_+),
    \end{equation}
where $\Kappa_0$ is the classical surface gravity at the horizon and $\xi$ is a dimensionless constant of the magnitude of the order $\hbar$ and has the structure
    \begin{equation} \label{eq:xi}
    \xi = \beta\frac{M_P^2}{M^2}.
    \end{equation}
Above $\beta$ is a numerical factor and $M_P$ is the Planck mass. The relation \eqref{eq:xi} is based on the relevant scale of mass (or length) of the problem which is the black hole mass $M$ compared to the Planck mass $M_P$. We assume that the terms of the order $\xi^2$ are negligible when compared to the terms of order $\xi$. Substituting equation \eqref{eq:xi} on the equation \eqref{eq:kappa_perturb_def}, we get

\begin{equation} \label{eq:kappa_back}
\Kappa = \Kappa_0(r_+)\left(1 + \frac{\sigma}{M^2(r_+)}\right),
\end{equation}
where $\sigma = \beta M_P^2$. The interpretation of the back reaction effect as a perturbation in surface gravity is useful for calculating the Hawking temperature due to the relationship between these two quantities given by \cite{Hawking} 
\begin{equation} \label{eq:T_H = kappa/2pi}
    T_H = \frac{\Kappa}{2\pi}.
\end{equation} 
Then it is possible to calculate the thermodynamic quantities. It is important to understand that the consideration of the back reaction effect performed here is known as \textit{1-loop} which means that the effect corresponds to the perturbative series dominant and following terms, as one can see by observing the relation \eqref{eq:kappa_perturb_def} \cite{Schweber, Mukhanov}. That is, the discussion presented here considers that the back reaction effect is related to the semiclassical contributions interpreted as conformal anomaly effects, in contrast to the approach of renormalization of the moment-energy tensor.

We first analyze the commutative spacetime. The line element presented by \eqref{eq:BS_metric} describes the behavior of the black string commutative spacetime and it can be used to calculate

\begin{equation} \label{eq:Kappa_0}
\Kappa_0(r_+) = \frac{3\alpha^2}{2}r_+.
\end{equation}
Substituting the equation \eqref{eq:Kappa_0} into the equation \eqref{eq:kappa_perturb_def}, we find

\begin{equation} \label{eq:kappa_back_comut}
\Kappa = \frac{3\alpha^2}{2}r_+\left(1 + \frac{16\sigma}{\alpha^6 r_+^6}\right),
\end{equation}
where we have used the relation $r_+ = \frac{(4M_+)^{\frac{1}{3}}}{\alpha}$ to write the last result in terms of $r_+$. Using the relation $T_H = \Kappa/2\pi$ we can write 

\begin{equation} \label{eq:T_B_comut}
\calT_H^{(c)} = \frac{3\alpha^2}{4\pi}r_+\left(1 + \frac{16\sigma}{\alpha^6 r_+^6}\right).
\end{equation}
Notice that the last equation reduces to the temperature without the presence of back reaction \eqref{eq:T_H_C} when one sets $\sigma = 0$, as expected. Therefore, we can write

\begin{equation} \label{eq:T_B_comut = T_H (1 +...)}
\calT_H^{(c)} = T_H^{(c)} \left(1 + \frac{16\sigma}{\alpha^6 r_+^6}\right).
\end{equation}

From the thermodynamical relation $dM = TdS$, the heat capacity can be obtained as

\begin{equation}
\calC^{(c)} = \frac{\calT_H^{(c)}\partial \calS^{(c)}}{\partial \calT_H^{(c)}} = \frac{\left(\frac{\partial M_+}{\partial r_+}\right)}{\left(\frac{\partial \calT_H^{(c)}}{\partial r_+}\right)}.
\end{equation}
Now using \eqref{eq:horizonte_BS}, we find

\begin{equation}
\frac{\partial M_+}{\partial r_+} = \frac{3}{4}\alpha^3 r_+^2,
\end{equation}
and using \eqref{eq:T_H_C}, \eqref{eq:T_B_comut}, we can write

\begin{eqnarray*}
\frac{\partial \calT_H^{(c)}}{\partial r_+} &=& \frac{3\alpha^2}{4\pi}\left[\left(1 + \frac{16\sigma}{\alpha^6 r_+^6}\right) + r_+\left(\frac{-96\sigma}{\alpha^6 r_+^7}\right)\right] \nonumber\\
%&=& \frac{3\alpha^2}{4\pi}\left(1 + \frac{16\sigma}{\alpha^6 r_+^6} - \frac{96\sigma}{\alpha^6 r_+^6}\right) \nonumber\\
&=& \frac{3\alpha^2}{4\pi}\left(1 - \frac{80\sigma}{\alpha^6 r_+^6}\right).
\end{eqnarray*}
This leads to 

\begin{equation} \label{eq:C_B_comut}
\calC^{(c)} = \frac{\pi\alpha r_+^2}{\left(1 - \frac{80\sigma}{\alpha^6 r_+^6}\right)}.
\end{equation}
If we compare the above result to \eqref{eq:C_+_comut}, we can write

\begin{equation} \label{eq:C_B_comut = C_+_comut(1 +..)}
\calC^{(c)} = \frac{C_+^{(c)}}{\left(1 - \frac{80\sigma}{\alpha^6 r_+^6}\right)}.
\end{equation}
The above equation reduces to \eqref{eq:C_+_comut} when one sets $\sigma = 0$, as expected.

\begin{figure}[h]
        \includegraphics[scale=0.5]{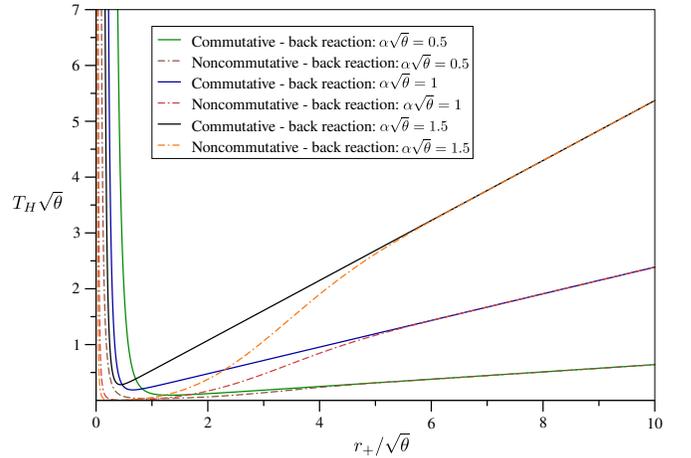}
        \caption{\label{fig:temp_B}\footnotesize Temperature in presence of back reaction for some values of $\alpha\sqrt{\theta}$. The horizontal axis is normalized by $\sqrt{\theta}$ and vertical axis is multiplied by this factor. The divergences occur in the not physical scale. We considered $\sigma^{-3}$ for this plot.}
    \end{figure}

Analyzing the back reaction effects on noncommutative spacetime, we obtain the surface gravity by considering the relation \eqref{eq:T_H = kappa/2pi} and the line element \eqref{eq:NC_BS_metric}. That is

\begin{equation}
    \Kappa_0 = \frac{3\alpha^2 r_+}{2} - \frac{\alpha^2r_+^4}{8\theta^{\frac{3}{2}}} \frac{e^{-\frac{r_+^2}{4\theta}}}{\gamma\left(\frac{3}{2},\frac{r_+^2}{4\theta}\right)},
\end{equation}
where the relation \eqref{eq:NC_BS_event horizon} has been employed. Then, writing the surface gravity in presence of back reaction

\begin{equation} \label{eq:Kappa_NC_m_theta}
    \Kappa = \frac{\alpha^2 r_+}{2}\left[3 - \frac{r_+^3}{4\theta^{\frac{3}{2}}}\frac{e^{-\frac{r_+^2}{4\theta}}}{\gamma\left(\frac{3}{2},\frac{r_+^2}{4\theta}\right)}\right]\left(1 + \frac{\sigma}{m_{\theta}^2(r_+)}\right).
\end{equation}
Now, we notice that it is possible to reach a simple relation between $m_\theta(r_+)$ and $r_+$ if one writes \eqref{eq:m_theta} modified by \eqref{eq:M_+} finds 

\begin{equation} \label{eq:m_teta(r_+)}
    m_\theta(r_+) = \frac{1}{4}\alpha^3r_+^3,
\end{equation}
substituting the equation \eqref{eq:m_teta(r_+)} into \eqref{eq:Kappa_NC_m_theta}, we find

\begin{equation} \label{eq:Kappa_NC}
\Kappa = \frac{\alpha^2 r_+}{2}\left[3 - \frac{r_+^3}{4\theta^{3/2}}\frac{e^{-\frac{r_+^2}{4\theta}}}{\gamma\left(\frac{3}{2},\frac{r_+^2}{4\theta}\right)}\right]\left(1 + \frac{16\sigma}{\alpha^6 r_+^6}\right),
\end{equation}
which leads to 

\begin{eqnarray} \label{eq:T_B_NC}
\calT_H &=& \frac{\alpha^2 r_+}{4\pi}\left[3 - \frac{r_+^3}{4\theta^{3/2}}\frac{e^{-\frac{r_+^2}{4\theta}}}{\gamma\left(\frac{3}{2},\frac{r_+^2}{4\theta}\right)}\right]\left(1 + \frac{16\sigma}{\alpha^6 r_+^6}\right) \nonumber\\
&=& T_H\left(1 + \frac{16\sigma}{\alpha^6 r_+^6}\right).
\end{eqnarray}
Considering the commutative limit $r_+/\sqrt{\theta}$ we obtain the equation \eqref{eq:T_B_comut}. The figure \ref{fig:temp_B} illustrates the results \eqref{eq:T_B_NC} and \eqref{eq:T_B_comut}.

 Analyzing the plot \ref{fig:temp_B} we notice that when the radius of the event horizon approaches to zero the presence of back reaction changes completely the behavior of temperature curves for both commutative and noncommutative cases. The behaviour of temperature in the presence of the back reaction increases dramatically when $r_+/\sqrt{\theta}$ approaches zero. Investigating the figure \ref{fig:temp_B} we see that the back reactions is relevant only for very short lengths, most specifically, at the range $\sqrt{\theta}$. We emphasize that in the noncommutative case we need to abandon the notion of geometric point. Therefore there are no distances smaller than the minimum length characteristic of the spacetime given by $\sqrt{\theta}$. In other words, lengths less than such value are meaningless, and therefore should not represent actual physical systems. Analyzing the plot \ref{fig:temp_B}, we notice that the back reaction effect is really significant in the range of distances less than the minimum length. That is, the Hawking temperature of the cylindrical black hole does not suffer changes in this description of gravity on a small scale since the scale less than $\sqrt{\theta}$ does not represent an acceptable range of distances.

Now, investigating the heat capacity we follow a similar procedure to performed in commutative spacetime case. For the noncommutative spacetime we find the heat capacity,

\begin{widetext}
    \begin{equation} \label{eq:C_B_NC}
\calC = \frac{\pi^{\frac{3}{2}}}{2}\alpha r_+^2\frac{\left(3 - \frac{\gamma'}{\gamma}r_+\right)}{\left[3\gamma - \gamma'' r_+^2 - 2\gamma'r_+ + \frac{{\gamma'}^2}{\gamma}r_+^2 - \frac{16\sigma}{\alpha^6 r_+^6}\left(15\gamma + \gamma'' r_+^2 - 4\gamma' r_+ - \frac{{\gamma'}^2}{\gamma}r_+^2\right)\right]}.
    \end{equation}
\end{widetext}
Notice that it is possible to rewrite the equation \eqref{eq:C_B_NC} in a simpler way in terms of $C_+$:

\begin{equation} \label{eq:C_B_NC = C_+(1/1+ ..}
\calC = \frac{C_+}{1 - \frac{16\sigma}{\alpha^6 r_+^6}\left[\frac{15\gamma + r_+\left(\gamma'' r_+ - 4\gamma' + \frac{{\gamma'}^2}{\gamma}r_+\right)}{3\gamma - r_+\left(\gamma'' r_+ + 2\gamma' - \frac{{\gamma'}^2}{\gamma}r_+\right)}\right]}.
\end{equation}
In the equation \eqref{eq:C_B_NC = C_+(1/1+ ..} $\gamma$ represents $\gamma\left(\frac{3}{2},\frac{r_+^2}{4\theta}\right)$ for simplicity. If we set $\sigma = 0$ into the relation \eqref{eq:C_B_NC = C_+(1/1+ ..} we recover the expected result given by equation \eqref{eq:C_NC}.

\begin{figure}[h]
        \includegraphics[scale=0.5]{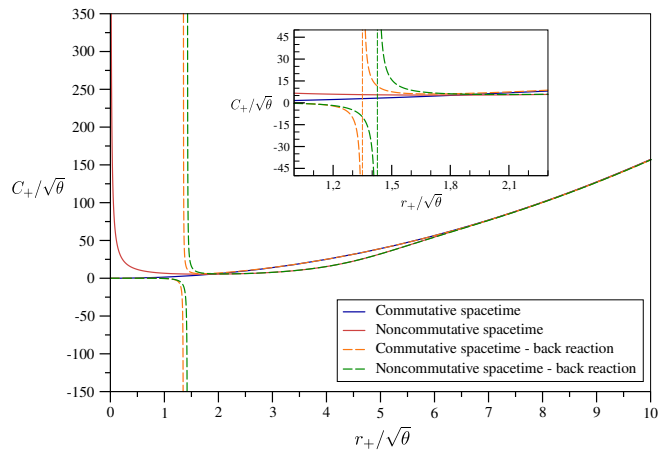}
        \caption{\label{fig:capa_B_0.5}\footnotesize Heat capacity in presence of back reaction. Here $\alpha\sqrt{\theta} = 0.5$ and $\sigma = 10^{-3}$.}
    \end{figure}

Unlike temperature, the heat capacity is influenced by the back reaction effect on the physical regime ($r_+ \geq \sqrt{\theta}$) and there are several implications. By investigating the commutative case (orange curve in figure \ref{fig:capa_B_0.5}), which is presented under the configuration $\alpha\sqrt{\theta} = 0.5$, we can point out that the vertical asymptote shown is located at $r_+ = 1.35\sqrt{\theta}$. This point divides the heat capacity plot into two regions: the region $r_+ < r_+ = 1.35\sqrt{\theta}$ is characterized by negative values. Such fact indicates thermodynamic instability and $C_+/\sqrt{\theta}$ continues to assume increasingly negative values then the curve approaches the asymptote at $1.35$. The region $r_+ > 1.35\sqrt{\theta}$ is characterized by positive values, which indicates thermodynamic stability in that region. Checking the point $r_ + = 1.35\sqrt{\theta}$ in figure \ref{fig:capa_B_0.5} we verify that it corresponds to a discontinuity because the heat capacity diverges on the right and on the left. It represents a point at which the behavior is modified (from instability on the left to stability on the right). We are, therefore, dealing with a black string thermodynamic phase transition. We have a first order phase transition as a consequence of the back reaction effect. The perturbation causes a symmetry breaking over the system, which corresponds to conformal symmetry breaking (conformal symmetry is that associated with conformal transformations which leave the metric tensor invariant at less than a scale factor. The generated group is the conformal group to which the Poincar?? group is a subgroup \cite{Fursaev, Lousto}). Therefore, the back reaction effect acts over the system causing a conformal anomaly which reflects quantum effects over the classical system. To summarize, the back reaction effect causes a thermodynamic phase transition when the system reaches determined temperature (or, equivalently, determined radius).

A more precise analysis is performed considering the temperature value corresponding to the phase transition \cite{Greiner,Mario}. The critical temperature can be found rewriting $r_+$ in terms of the Hawking temperature. On another hand, the phase transition can be identified by the heat capacity (or specific heat) versus absolute temperature analysis. This can be verified in figure \ref{fig:capa_B_0.5} where we have the heat capacity versus radius of the event horizon $r_+$. Moreover $r_+$ plays the role of temperature as control parameter in figure \ref{fig:capa_B_0.5}.  Knowing the critical value of radius which the heat capacity diverges, we can calculate the critical temperature values. We can perform such analysis by regarding the dashed green curve which describes the noncommutative spacetime case: we have a vertical line at the point $r_+ = 1.426 \sqrt{\theta}$; this point divides the plot into two regions: the region on the the left presents thermodynamic instability, and the region of thermodynamic stability on the right. Following the same analysis that we have just done, this point shows the position (with the corresponding critical temperature) where the phase transition occurs. Similar considerations can be made regarding the dashed orange curve which describes the commutative spacetime case. To calculate the critical temperature for the commutative and noncommutative spacetime cases, we just replace the critical radius values, $r_+^{(c)}/\sqrt{\theta} = r_{\mbox{\footnotesize critical}}^{(c)} = 1.35$ for the commutative case and $r_+/\sqrt{\theta} = r_{\mbox{\footnotesize critical}} = 1.426$ for the noncommutative spacetime case, in the respective temperature expressions \eqref{eq:T_B_comut} and \eqref{eq:T_B_NC}. Then we obtain the following values:

\begin{equation} \label{eq:T_critico_0.5}
\begin{aligned}
	\calT_{\mbox{\footnotesize c}}^{(c)}\sqrt{\theta} &= 0.0136 \\
	\calT_{\mbox{\footnotesize c}}\sqrt{\theta} &= 0.0183,
\end{aligned}
\qquad \alpha\sqrt{\theta} = 0.5.
\end{equation}

We notice that the critical temperature is not affected expressively under the choice of parameters in the noncomutativity of spacetime case. In this case, the difference in the critical temperature $\calT_{\mbox{\footnotesize c}}$ in equation \eqref{eq:T_critico_0.5} is noteworthy only in the third decimal place $0.0047$. This shows that the phase transition temperature is not significantly modified when the black string radius considered is slightly different. Still on the analysis of the plot \ref{fig:capa_B_0.5}, we notice that the noncommutativity of spacetime causes the vertical asymptote to present a shift about the order of $0.076$ considering the change from point $1.350$ to point $1.426$. It means that the phase transition occurs at a higher value of black string radius due to the noncommutative effects. Notice also that the perfect overlap of the curves elaborated with the consideration of the back reaction effect (green for the noncommutative case and orange for the commutative case) for values higher than the critic point of the event horizon indicates that results in the equations \eqref{eq:C_B_NC} and \eqref{eq:C_B_comut} are correct and they recover the previous results when $r_+/\sqrt{\theta}$ is large, as expected.

\begin{figure}[h]
        \includegraphics[scale=0.51]{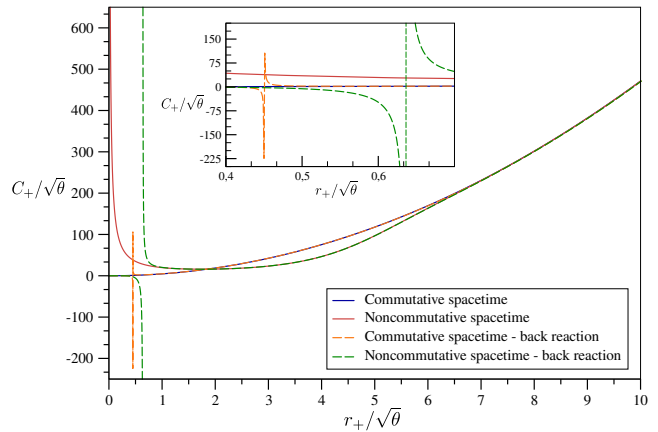}
        \caption{\label{fig:capa_B_1.5}\footnotesize Heat capacity in presence of back reaction. Here $\alpha\sqrt{\theta} = 1.5$ and $\sigma = 10^{-3}$.}
    \end{figure}

Now investigating the plot presented in the figure \ref{fig:capa_B_1.5} we find an expressive modification of the curve behaviour regarding the heat capacity in the presence of back reaction in commutative spacetime. When $\alpha\sqrt{\theta} = 1.5$ the system presents a finite discontinuity in the curve (dashed orange line), in contrast to the infinite discontinuity that occurs in the figure \ref{fig:capa_B_0.5} in the case $\alpha\sqrt{\theta} = 0.5$. To summarize, in commutative spacetime the system presents a second order phase transition due to the presence of back reaction, meanwhile in noncommutative spacetime the system presents a first order phase transition due back reaction. Following the same procedure previously performed, we determine the respective critical temperatures associated to the commutative and noncommutative spacetimes. We rewrite the critical radius $r_{\mbox{\footnotesize critical}}$ in terms of the critical Hawking temperature and then plot the subsequent relation in order to verify the desired critical value. For both cases, the critical temperatures are given by the following values

\begin{equation} \label{eq:T_critico_1.5}
\begin{aligned}
	\calT_{\mbox{\footnotesize c}}^{(c)}\sqrt{\theta} &= 0.2826 \\
	\calT_{\mbox{\footnotesize c}}\sqrt{\theta} &= 0.0139,
\end{aligned}
\qquad \alpha\sqrt{\theta} = 1.5.
\end{equation}

Immediately we can observe that for the commutative case the temperature value is about an order of magnitude greater than the value associated with the noncommutative spacetime. It is also an order of magnitude greater than the value associated with the other cases under the configuration $\alpha\sqrt{\theta} = 0.5$. This fact reveals a difference about the order $0.269$ between the temperature values in \eqref{eq:T_critico_1.5} and \eqref{eq:T_critico_0.5}. Therefore, we realize that the system is sensitive to the choice of the parameter associated to the cosmological constant.  The cosmological constant plays an important role in the interpretation of the system because its magnitude can be associated with the density of vacuum energy, according to general relativity. The sensitivity of the system to the constant $\alpha\sqrt{\theta}$ (i.e. $\alpha\sqrt{\theta} = 0.5$ or $\alpha\sqrt{\theta} = 1.5$) reveals the influence of vacuum energy on the spacetime of the black string. 

Considering the cosmological constant ($\Lambda$) as a parameter for the accelerated expansion of the universe, we understand that the thermodynamic properties depends on the acceleration of the universe expansion caused by vacuum energy. Moreover, the dependency on the cosmological constant ($\Lambda$) is expressed in terms of the minimum length of spacetime $\sqrt{{\theta}}$, a candidate for fundamental length scale \cite{Szabo}. In other words, when vacuum energy density is written as a semi-integer multiple of the noncommutativity parameter $\alpha\sqrt{\theta} = 0.5$ (because $\alpha^2 = - \Lambda/3$), the cylindrical black hole presents a first order phase transition. On another hand, when the vacuum energy density is written as a larger multiple $\alpha\sqrt{\theta} = 1.5$ the predicted behavior in commutative spacetime case is a second order phase transition.

%%%%%%%%% %%%%%%%%% %%%%%%%%% %%%%%%%%% %%%%%%%%% %%%%%%%%% %%%%%%%%% %%%%%%%%% %%%%%%%%% %%%%%%%%%
\section{Conclusions}
%%%%%%%%% %%%%%%%%% %%%%%%%%% %%%%%%%%% %%%%%%%%% %%%%%%%%% %%%%%%%%% %%%%%%%%% %%%%%%%%% %%%%%%%%%

In order to summarize our discussion, we remark that the black string noncommutative spacetime is a prediction about the behavior of spacetime when one considers the cylindrical black hole solution \eqref{eq:BS_metric} in short scale, about the Planck scale order. The relation between noncommutative geometry and the theory of black strings leads to the line element \eqref{eq:NC_BS_metric} which represents the noncommutative spacetime of a black string \cite{Singh}. Such fact in addition to the fact that it is possible to associate a well defined temperature to a black hole motivates the investigation of the thermodynamic properties of black strings in short scale.

The black string thermodynamics presents modifications in short scale according to the equation \eqref{eq:T_H_NC}. The temperature has no longer the linear behavior being less than the commutative prediction. Investigating the entropy in equation \eqref{eq:entropy}, the Bekenstein area law is not observed in noncommutative spacetime case. This fact is due to the noncommutativity effects because when we take the commutative spacetime limit  ($r_+/\sqrt{\theta} \rightarrow \infty$) the Bekenstein area law is recovered. In a preliminary analysis, the black string can be considered a stable thermodynamic system because the heat capacity is always positive. The positive behavior of the heat capacity is an affirmation that the black string shall not face any phase transition. This scenario changes significantly if one considers perturbative effects. The predictions, considering perturbative effects, are quite interesting when we analyze the system from the perspective of phase transitions. An analysis of the heat capacity behavior in figures \ref{fig:capa_B_0.5} and \ref{fig:capa_B_1.5} shows us that the black string faces phase transitions in the presence of back reaction. Moreover, the order of the phase transitions depends on the parameter $\alpha\sqrt{\theta}$ which involves the cosmological constant and the minimum length of spacetime, as we discussed in section \ref{sec:back}. Thus the phase transition order depends on the relation between the acceleration of expansion of the universe and the minimum length of the spacetime. It depends on the relation between the density of vacuum energy and the minimum length associated with spacetime. This result can be important for analysis of conformal symmetry breaking in black strings.

Investigating the free energy of the system, we notice that work can be done over or by the black string as long as the appropriate conditions are established and a temperature is determined.

%%%%%%%%% %%%%%%%%% %%%%%%%%% 
\begin{acknowledgments}
The authors thank to Brazilian agency CAPES - Coordena\c{c}\~ao de Aperfei\c{c}oamento de Pessoal de N\'{i}vel Superior (Coordination for the Improvement of Higher Education Personnel) for financial support.
\end{acknowledgments}
%%%%%%%%% %%%%%%%%% %%%%%%%%%

%%%%%%%%% %%%%%%%%% %%%%%%%%% %%%%%%%%% %%%%%%%%% %%%%%%%%% %%%%%%%%% %%%%%%%%% %%%%%%%%% %%%%%%%%%
\appendix*
\section{\label{sec_ap:gamma} INCOMPLETE LOWER GAMMA FUNCTION}
%%%%%%%%% %%%%%%%%% %%%%%%%%% %%%%%%%%% %%%%%%%%% %%%%%%%%% %%%%%%%%% %%%%%%%%% %%%%%%%%% %%%%%%%%%

In this section, we discuss some aspects of the incomplete lower gamma function. The investigation of the noncommutative effects influence on the thermodynamics of black holes demonstrates that such function plays a crucial role in the mathematical description of these systems. This function is important whether the black hole is spherical \cite{Nicolini1,Nicolini2} or cylindrical \cite{Singh}.

%%%%%%%%% %%%%%%%%% %%%%%%%%%
\subsection{\label{subsec_ap:def_gama} Definition}
%%%%%%%%% %%%%%%%%% %%%%%%%%%

The incomplete lower gamma function is defined by

\begin{equation} \label{eq:def_gama}
\gamma\left(s,x\right) = \int_0^x t^{s-1}e^{-t} dt.
\end{equation}
The definition \eqref{eq:def_gama} above clarifies the name of such function: it is the integral representation of the complete gamma function except for the finite upper index in the integral. In a similar way, one can define the incomplete upper gamma function as

\begin{equation} \label{eq:def_gama_sup}
\Gamma\left(s,x\right) = \int_x^{\infty} t^{s-1}e^{-t}dt
\end{equation}
in such a way that

\begin{equation} \label{eq:Gama_total}
\Gamma(s) = \gamma\left(s,x\right) + \Gamma\left(s,x\right),
\end{equation}
for all $s > 0$ and $x \geq 0$. The function \eqref{eq:def_gama_sup} is also called as complementary incomplete gamma function which were investigated by Prym in 1877 \cite{Jameson}. It is important to remark that the integral \eqref{eq:def_gama} converges for $s > 0$ and the analysis for $s < 0$ can be done, but it is not important here. 

The functions \eqref{eq:def_gama}, \eqref{eq:def_gama_sup} and \eqref{eq:Gama_total} are functions of $x$ and have the fixed parameter $s$ in such a way that if one considers $x \rightarrow \infty$, the incomplete lower gamma function can be written as \cite{Abramowitz}

\begin{equation} \label{eq:expan_gama_1}
\gamma\left(s,\infty\right) = \int_0^\infty t^{s-1}e^{-t}dt = \Gamma(s).
\end{equation}
This property is one of the most important properties for our purposes and now we shall study some others.

%%%%%%%%% %%%%%%%%% %%%%%%%%%
\subsection{\label{subsec_ap:propriedades} Properties}
%%%%%%%%% %%%%%%%%% %%%%%%%%%

We remark some useful properties:

\begin{itemize}
\item \textbf{Series expansion}
\end{itemize}

\begin{eqnarray} \label{eq:prop_expansao_gama}
\gamma\left(s,x\right) &=& \int_0^x t^{s-1}e^{-t}dt \nonumber\\
&=& \int_0^x t^{s-1} \sum_{k=0}^{\infty} (-1)^k \frac{t^k}{k!} dt \nonumber\\
&=& \sum_{k=0}^{\infty} \frac{(-1)^k}{k!} \int_0^x t^{s + k - 1} dt \nonumber\\
&=& \sum_{k=0}^{\infty} (-1)^k\frac{x^{s + k}}{(s + k)k!}.
\end{eqnarray}

\begin{itemize}
\item \textbf{Derivative}
\end{itemize}

\begin{equation} \label{eq:prop_deriva_gama}
\frac{\partial \gamma\left(s,x\right)}{\partial x} = x^{s-1}e^{-x}.
\end{equation}
This relation can be proven by using the previous property \eqref{eq:prop_expansao_gama} as follows

\begin{eqnarray}
\frac{\partial}{\partial x} \sum_{k=0}^{\infty} (-1)^k\frac{x^{s + k}}{(s + k)k!} 
%= \sum_{k=0}^{\infty} \frac{(-1)^k}{(s + k)k!}\frac{\partial x^{s + k}}{\partial x} \nonumber\\
&=& x^{s-1}\sum_{k=0}^{\infty} (-1)^k\frac{x^k}{k!} \nonumber\\
&=& x^{s-1}e^{-x}.
\end{eqnarray}
It can be proven that \cite{Abramowitz}

\begin{equation}
    \frac{\partial \gamma\left(s,x\right)}{\partial x} = - \frac{\partial \Gamma\left(s,x\right)}{\partial x} 
\end{equation}

\begin{itemize}
\item \textbf{Asymptotic expansion}
\end{itemize}

We can find an incomplete lower gamma function Taylor series by performing the substitution $t = u + x$ into integral \eqref{eq:def_gama_sup} as

\begin{eqnarray}
\Gamma(n + 1, x) 
%&=& \int_x^\infty t^n e^{-t} dt \nonumber\\
&=& e^{-x}\int_0^\infty (u + x)^n e^{-u} du \nonumber\\
&=& e^{-x}\int_0^\infty \sum_{k=0}^\infty \binom{n}{p} u^{n-p} x^p e^{-u} du,
\end{eqnarray}
where $p = n - k$ and we used $\binom{n}{k} = \binom{n}{n-k}$. We find

\begin{eqnarray} \label{eq:expan_gama_sup}
\Gamma(n + 1, x) &=&  e^{-x}\sum_{k=0}^\infty \binom{n}{p} x^p\int_0^\infty u^{n-p} e^{-u} du \nonumber\\
&=& e^{-x}\sum_{k=0}^\infty \binom{n}{p} x^p \Gamma(n - p + 1) \nonumber\\
%&=& e^{-x}\sum_{k=0}^\infty \frac{n!}{p!(n - p)!}x^p (n - p)! \nonumber\\
&=& e^{-x}\sum_{k=0}^\infty \frac{\Gamma(n + 1)}{\Gamma(n - k + 1)} x^{n - k}.
\end{eqnarray}

Now, if one realizes that $\gamma(n+1,x) + \Gamma(n+1,x) = \Gamma(n+1)$, by \eqref{eq:Gama_total}, and uses \eqref{eq:expan_gama_sup} one finds 

\begin{eqnarray*}  \label{eq:expan_gama_inf_gama_sup}
\gamma\left(n+1,x\right) &=& \Gamma\left(n+1\right) - \Gamma\left(n+1,x\right) \\
&=& \Gamma(n+1) - e^{-x}\sum_{k=0}^\infty \frac{\Gamma(n + 1)}{\Gamma(n + 1 - k)} x^{n - k}.
\end{eqnarray*}
For our purposes, section \ref{sec:NC_spacetime}, $(n + 1 = s = 3/2)$ and for large $x$, we have

\begin{eqnarray} \label{eq:expan_assin_gama}
\gamma\left(\frac{3}{2},x\right) &=& \Gamma\left(\frac{3}{2}\right) - \Gamma\left(\frac{3}{2},x\right) \nonumber\\
&=& \frac{\sqrt{\pi}}{2}\left[1 - e^{-x}\sum_{k=0}^{\infty} \frac{x^{(1-2k)/2}}{\Gamma\left(\frac{3}{2} - k\right)}\right] \nonumber\\
&\approx & \frac{\sqrt{\pi}}{2} - \sqrt{x}e^{-x} - \cdots
\end{eqnarray}
Here we made $k = 0$, the first iteration. The equation \eqref{eq:expan_assin_gama} can be seen as a demonstration for the equation \eqref{eq:expan_gama_1} in terms of representation of functions by series. When $x$ is large, the first correction, the term $\sqrt{x}e^{-x}$ in equation \eqref{eq:expan_assin_gama} above goes to zero and we recover $\Gamma\left(\frac{3}{2}\right) = \frac{\sqrt{\pi}}{2}$ as it is expressed in equation \eqref{eq:expan_gama_1}. This property is used all over this work.

\begin{itemize}
\item \textbf{Useful expressions}
\end{itemize}

\begin{equation} \label{eq:expres_uteis_func_gama_1}
    \gamma(1,x) = 1 - e^{-x}
\end{equation}

\begin{equation} \label{eq:expres_uteis_func_gama_2}
\gamma\left(\frac{3}{2},x\right) = - \sqrt{x}e^{-x} + \frac{1}{2}\gamma\left(\frac{1}{2},x\right) 
\end{equation}

\begin{equation} \label{eq:func_erro}
    \gamma\left(\frac{1}{2},x\right) = \sqrt{\pi}\erf(\sqrt{x})
\end{equation}
where

\begin{equation}
\erf(z) = \int_0^z \frac{2e^{-t^2}}{\sqrt{\pi}} dt
\end{equation}
is the \textit{error function}.

\begin{eqnarray}
    I &=& \int_a^b e^{-\lambda x^2} dx \nonumber\\
    &=& \frac{1}{2\lambda^{\frac{1}{2}}}\left[\sqrt{\pi} - \Gamma\left(\frac{1}{2}, \lambda b^2\right) - \gamma\left(\frac{1}{2}, \lambda a^2\right)\right]
\end{eqnarray}

In the context of this work, it is also useful to write 

\begin{equation}
    \frac{d}{dr_+} \gamma\left(\frac{3}{2},\frac{r_+^2}{4\theta}\right) = \frac{r_+^2}{4\theta^{\frac{3}{2}}}e^{\frac{r_+^2}{4\theta}}
\end{equation}

%%%%%%%%% %%%%%%%%% %%%%%%%%%
%\nocite{*}
\bibliography{Bibliography} 
%%%%%%%%% %%%%%%%%% %%%%%%%%%

%%%%%%%%% %%%%%%%%% %%%%%%%%% %%%%%%%%% %%%%%%%%% %%%%%%%%% %%%%%%%%% %%%%%%%%% %%%%%%%%% %%%%%%%%%
\end{document}